\documentclass[preprint,floats,aps,prl,eqnum,showpacs,column]{revtex4}
\usepackage{color,graphicx}
\usepackage{amsfonts}
\usepackage{amssymb}
\usepackage{ulem}

\begin{document}

\title{Signed magnification sums for general spherical lenses}

\author{Naoki Tsukamoto\footnote{Electronic address:11ra001t@rikkyo.ac.jp}
and Tomohiro Harada
}

\affiliation{
Department of Physics, Rikkyo University, Tokyo 171-8501, Japan 
}
\date{\today}

\begin{abstract}
It is well known that the sum of signed magnifications is invariant for mass lens systems. 
In this paper, we discuss the signed magnification sums of general spherical lens models 
including the singular isothermal sphere, the Schwarzschild lens and the Ellis wormhole, 
the last of which is an example of the traversable wormholes of the Morris-Thorne class. 
We show that the signed magnification sums are a very useful tool to distinguish exotic lens objects. 
For example, we show that one can distinguish the Ellis wormholes from the Schwarzschild lens with the signed magnification sums.
\end{abstract}


\preprint{
RUP-12-11
}

\maketitle

\section{I. ~Introduction}
Gravitational lensing is not only a useful tool for astrophysics and cosmology 
(see Schneider \textit{et al.} \cite{Gravitational_lenses}, Perlick \cite{Perlick_2004_Living_Rev,Perlick_2010} and references therein 
for the details of the gravitational lens) 
but also is an interesting topic in the field of mathematical physics.

It is well known that the sum of signed magnifications  
is invariant in the weak field limit for mass lens systems.
Witt and Mao investigated the magnifications for lensing by double lenses and found that the signed magnification sums of the five images become unity inside a caustic \cite{Witt_Mao_1995}.
Rhie gave another proof for the invariance
of the signed magnification sums of gravitational lensing by double lenses and applied it to the n-point lens systems \cite{Rhie_1997}.
The signed magnification sums of the simple galaxy models which are variations on the singular isothermal sphere were studied by Dalal \cite{Dalal_1998},
and those of quadruple lenses were investigated by Witt and Mao \cite{Witt_Mao_2000}. 
Dalal and Rabin showed that residue integrals provide a simple
explanation for the invariance of the signed magnification sums  \cite{Dalal_Rabin_2001}.
Recently, Werner showed that the signed magnification invariant is a topological invariant \cite{Werner_2007}.
The local magnification relations with a subset of the total number of lensed images have been investigated eagerly 
\cite{Werner_2009,Aazami_Petters_2009_Mar,Aazami_Petters_2009_Aug,Aazami_Petters_2010,Petters_Werner_2010,Aazami_Petters_Rabin_2011}.

Since gravitational lensing was predicted about one hundred years ago, 
mass lens systems have been mainly investigated. 
However, curved spacetimes such as wormhole spacetimes also cause gravitational lens effects
(see Visser \cite{Lorentzian_Wormholes} for the details of the wormholes).
Since gravitational lensing of the wormholes was pioneered by Kim and Cho \cite{Kim_Cho_1994} and Cramer \textit{et al.} \cite{Cramer_Forward_Morris_Visser_Benford_Landis_1995},
many interesting aspects of gravitational lensing by various wormholes have been investigated 
\cite{Nandi_Zhang_Zakharov_2006,Safonova_Torres_Romero_2001,Eiroa_Romero_Torres_2001,Safonova_Torres_Romero_2002,Safonova_Torres_2002,Rahaman_Kalam_Chakraborty_2007,Dey_Sen_2008}.

The Ellis spacetime~\cite{Ellis_1973} 
is an example of static, spherically symmetric traversable wormholes.
Chetouani and Clement derived 
the deflection angle of light on it
and calculated the scattering cross-section \cite{Chetouani_Clement_1984}. 
Perlick investigated the gravitational lensing effects of the light ray through the Ellis wormhole throat by using the full lens equation \cite{Perlick_2004_Phys_Rev_D} 
and Nandi \textit{et al.} \cite{Nandi_Zhang_Zakharov_2006} applied the analysis of the strong field limit \cite{Virbhadra_Ellis_2000,Bozza_Capozziello_Iovane_Scarpetta_2001,Virbhadra_Ellis_2002,Bozza_2002}. 

It was pointed out that 
the qualitative features of gravitational lensing in the Ellis spacetime are similar to the ones in the Schwarzschild spacetime 
\cite{Perlick_2004_Phys_Rev_D,Nandi_Zhang_Zakharov_2006,Tejeiro_Larranaga_2012}. 
However, Abe showed that one can distinguish between the Ellis wormholes and mass lens objects with their light curves in the weak field limit \cite{Abe_2010}.
The Ellis wormholes could be detected with the astrometric image centroid trajectory in the weak field limit  \cite{Toki_Kitamura_Asada_Abe_2011}  
and with the Einstein ring and the relativistic Einstein rings \cite{Tsukamoto_Harada_Yajima_2012}.
Recently, Nakajima and Asada \cite{Nakajima_Asada_2012} recalculated the deflection angle of light on the Ellis spacetime 
and proved that Dey and Sen \cite{Dey_Sen_2008}'s calculation is only
correct at the lowest order in the weak field limit, 
while the conclusions by Abe \cite{Abe_2010} and Toki \textit{et al.}
\cite{Toki_Kitamura_Asada_Abe_2011} are still valid. 

In this paper, we will show that 
the signed magnification sum would be a powerful tool to research the lens objects as well as the total magnification and the magnification ratio 
if we observe a multiple image.
In particular, we will show that one can distinguish between the Ellis wormhole lens and the Schwarzschild lens with the signed magnification sums. 
We may test the hypotheses of the astrophysical wormholes 
\cite{Harko_Kovacs_Lobo_2009,Abdujabbarov_Ahmedov_2009,Pozanenko_Shatskiy_2010,Kardashev_Novikov_Shatskiy_2007} 
with the gravitational lensing in the future.

This paper is organized as follows.
In Sec. II we will discuss the signed magnification sums of the general spherical lens models 
including the singular isothermal sphere, the Schwarzschild lens and the Ellis wormhole. 
We will number the real solutions of the lens equation
because the signed magnification sums are physical invariants only when all the solutions are real.    
In Sec. III we discuss the signed magnification sums of the general spherical lens model in the directly aligned limit 
and we show that one can distinguish the general spherical lenses.
In Sec. IV we briefly review the Ellis wormhole spacetime and the deflection angle of light on it.
In Sec. V we summarize and discuss our result.
In this paper we use the units in which the light speed $c=1$.

\section{II. ~The signed magnification sums of the general spherical lens}
It is well known that the sum of signed magnifications is invariant for mass lens systems in the maximal-image domains. 
In this section, we will calculate the signed magnification sum of the general spherical lens model
and count the number of the images.

Now we will consider the case where both the observer and the source object are far from the lensing object
or $D_{l}\gg b$ and $D_{ls}\gg b$, 
where $b$, $D_{l}$ and $D_{ls}$ are the impact parameter of the photon 
and the separations between the observer and the lens and between the lens and the source, respectively.
The configuration of gravitational lensing is given in Fig. 1.
\begin{figure}[htbp]
\begin{center}
\includegraphics[width=80mm]{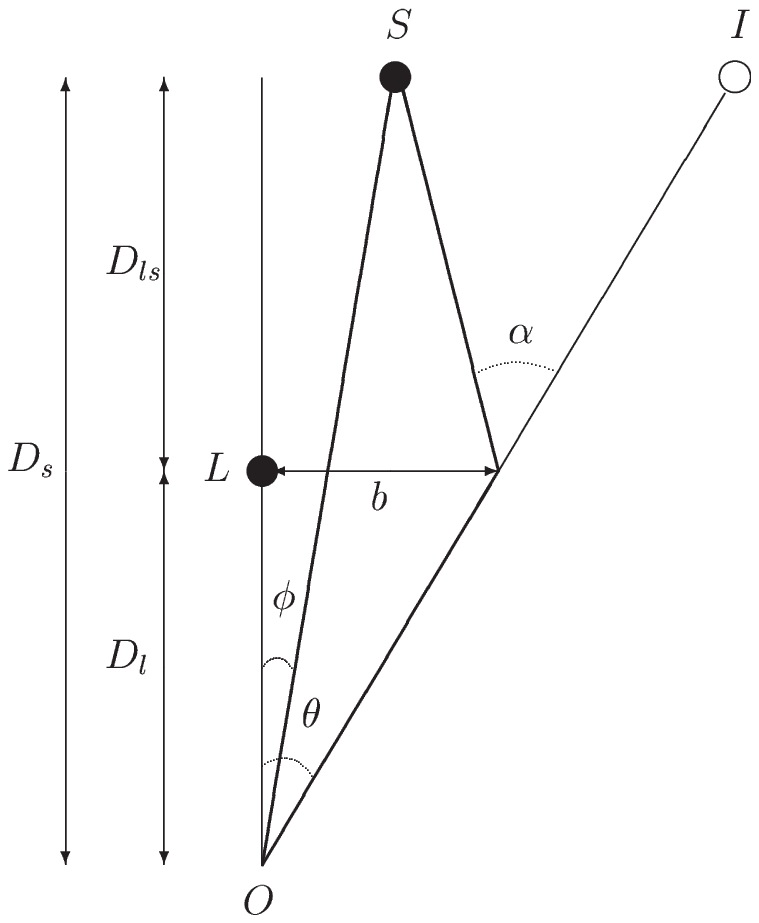}
\end{center}
\caption{The configuration of gravitational lensing.
The light rays emitted by the source $S$ are deflected by the lens $L$ 
and reach the observer $O$ with the lensed image angle $\theta$, instead of the source angle $\phi$. 
$b$ and $\alpha$ are the impact parameter of the photon and the deflection angle, respectively. 
$D_{l}$ and $D_{ls}$ are the separations between the observer and 
the lens and between the lens and the source, respectively.
$D_{s}=D_{l}+D_{ls}$ is the separation between the observer and the source.
}
\end{figure}
Then, the lens equation is given by 
\begin{equation}\label{eq:lens_equation}
D_{ls}\alpha=D_{s}(\theta-\phi),
\end{equation}
where $\alpha$ is the deflection angle, 
$\theta$ and $\phi$ are the image angle and
the source angle from the observer, respectively, 
and $D_{s}=D_{l}+D_{ls}$ is the separation between the observer and the source.
Note that we have assumed $|\alpha|\ll 1$, $|\theta|\ll 1$ and $|\phi|\ll 1$.

We consider the general spherical lens model with the deflection angle, parametrized by
\begin{eqnarray}
\alpha=\pm Cb^{-n}=\pm \frac{C}{D_{l}^{n}}\theta^{-n},
\end{eqnarray}
where $C$ is a positive constant and $n$ is a non-negative integer and we have used the relation $b=D_{l}\theta$.
If $n$ is odd, then the sign is only the upper one, while if $n$ is even, then the sign is the upper one for $\theta > 0$ and the lower one for $\theta < 0$. 
Thus, we have to treat two lens equations when $n$ is even.
This lens model describes the singular isothermal sphere, the Schwarzschild lens and the Ellis wormhole for $n=0$, $1$ and $2$, respectively.
The case where $n \geq 3$ would describe some exotic lens objects and the gravitational lens effect of modified gravitational theories.
The following discussion does not depend on the value of $C$.

The lens equation is given by
\begin{eqnarray}\label{eq:n+1th_lens_eq}
\hat{\theta}^{n+1}-\hat{\phi}\hat{\theta}^{n}\mp 1=0,
\end{eqnarray}
where
\begin{eqnarray}
\hat{\theta}\equiv \frac{\theta}{\theta_{0}}  \quad  \mathrm{and} \quad  \hat{\phi}\equiv \frac{\phi}{\theta_{0}},
\end{eqnarray}
and 
\begin{eqnarray}
\theta_{0} \equiv \left( \frac{D_{ls}C}{D_{s}D_{l}^{n}} \right)^{\frac{1}{n+1}}
\end{eqnarray}
is the Einstein ring angle.
We can concentrate ourselves on the case where the source angle $\phi$ is positive for symmetry.
The solutions $\hat{\theta}_{1}$, $\hat{\theta}_{2}$, $\cdots$, $\hat{\theta}_{n+1}$ of the lens equation (\ref{eq:n+1th_lens_eq}) of $(n+1)$-th degree satisfy
\begin{eqnarray}\label{eq:n+1th_solution}
\prod^{n+1}_{i=1} ( \hat{\theta}-\hat{\theta}_{i})=0.
\end{eqnarray}
For $n\geq 1$ we compare Eq. (\ref{eq:n+1th_lens_eq}) with Eq. (\ref{eq:n+1th_solution}) and obtain 
\begin{eqnarray}
\sum^{n+1}_{i=1}\hat{\theta}_{i}=\hat{\phi}, 
\end{eqnarray}
and
\begin{eqnarray}
\sum_{i<j}\hat{\theta}_{i}\hat{\theta}_{j}=-\delta_{1n},
\end{eqnarray}
where $\delta_{1n}=0$ for $n \geq 2$ and $\delta_{1n}=1$ for $n=1$.
Using the both equations, we obtain 
\begin{eqnarray}
\hat{\phi}^{2}
&=&\left( \sum^{n+1}_{i=1}\hat{\theta}_{i} \right)^{2} \nonumber\\ 
&=&\sum^{n+1}_{i=1}\hat{\theta}_{i}^{2}-2\delta_{1n}.
\end{eqnarray}
This implies 
\begin{eqnarray}\label{eq:global_invariant}
\sum^{n+1}_{i=1}\frac{\hat{\theta}_{i}}{\hat{\phi}}\frac{d\hat{\theta}_{i}}{d\hat{\phi}}=1.
\end{eqnarray}
Note that these solutions $\hat{\theta}_{i}$ may be complex 
and not all the magnifications are always physical 
and that Eq. (\ref{eq:global_invariant}) is satisfied regardless of the sign of Eq. (\ref{eq:n+1th_lens_eq}).

Now we will count the number of the images.
We can express the lens equation (\ref{eq:n+1th_lens_eq}) as follows 
\begin{eqnarray}\label{eq:n+1th_lens_eq_apart_form}
\pm \hat{\theta}^{-n}= \hat{\theta} - \hat{\phi}.
\end{eqnarray}
In the following, we will make 
analysis for cases (i) where $n$ is odd, (ii)
where $n$ is even and positive, and (iii) $n=0$, separately.

\subsection{(i) $n$ is odd.}
In the case where $n$ is odd, the lens equation is given by 
\begin{eqnarray}\label{eq:lens_odd}
\hat{\theta}^{-n}= \hat{\theta} - \hat{\phi}.
\end{eqnarray}
The solutions are given by the intersections of $y=1/x^{n}$ and $y=x-\hat{\phi}$.
Figure 2 shows the left-hand side and the right-hand side of the lens equation and the intersections for $n=1$.
\begin{figure}[htbp]
 \begin{center}
 \includegraphics[width=80mm]{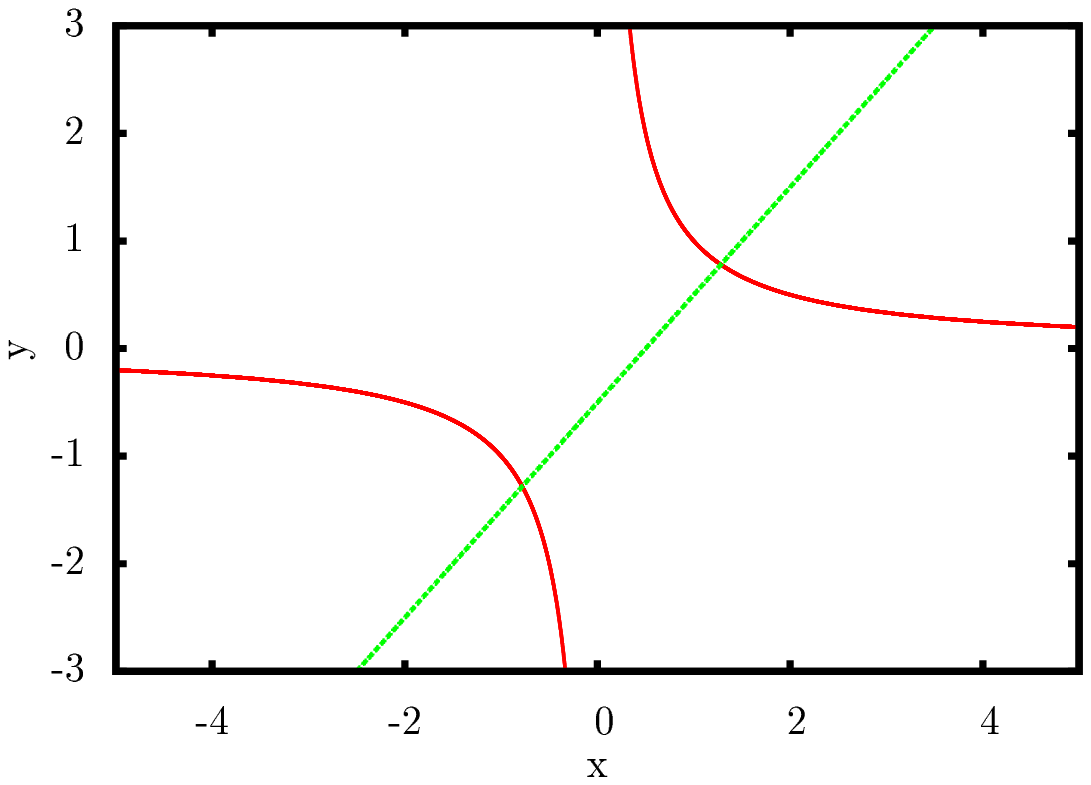}
 \end{center}
 \caption{The solid (red) lines $y=1/x^{n}$ and the broken (green) line $y=x-\hat{\phi}$, respectively, 
 correspond to the left-hand side and the right-hand side of the lens equation (\ref{eq:lens_odd}) for $n=1$ and $\hat{\phi}=0.5$.
 The intersections correspond to the real solutions of the lens equation.}
 \label{fig:three}
\end{figure}
We find a positive solution $\theta_{+}$ and a negative solution
$\theta_{-}$ regardless of the value for $\hat{\phi}$.
We also can see the positive solution $\hat{\theta}_{+} \sim \hat{\phi}$ and the negative solution $\hat{\theta}_{-} \sim 0$ for $\hat{\phi} \gg 1$.
We also can see that $n=1$ is the only case where all the solutions of the lens equation are real 
and the physical singed magnification sum is always unity.
The lens with $n=1$ and $C=4GM$ is the Schwarzschild lens, 
where $G$ is Newton's constant and $M$ is the lens mass.
Thus, its signed magnification sum is always unity.

\subsection{(ii) $n\geq 2$ is even.}
We consider the case where $n\geq 2$ is even. 
The lens equation is obtained by 
\begin{eqnarray}
\pm \hat{\theta}^{-n}= \hat{\theta} - \hat{\phi}.
\end{eqnarray}
The solutions are given by intersections of $y=1/x^{n}$ for $x>0$ and $y=-1/x^{n}$ for $x<0$ and $y=x-\hat{\phi}$.
This gives a figure which is very similar to Fig. 2
and we obtain a positive solution $\theta_{+}$ and a negative solution
$\theta_{-}$ regardless of the value for $\hat{\phi}$.
The signed magnification sum (\ref{eq:global_invariant}) is not a physical quantity 
because it includes one or more non-real solutions in this case.

\subsection{(iii) $n=0$.}
For $n=0$, the lens equation is given by
\begin{eqnarray}\label{eq:lens_0}
\pm 1= \hat{\theta} - \hat{\phi}.
\end{eqnarray}
The solutions are given by one or two intersections of $y=1$ for $x>0$ and $y=-1$ for $x<0$ and $y=x-\hat{\phi}$.
Figure 3 shows the left-hand side and the right-hand side of the lens equations and the intersections.
\begin{figure}[htbp]
\begin{center}
\includegraphics[width=80mm]{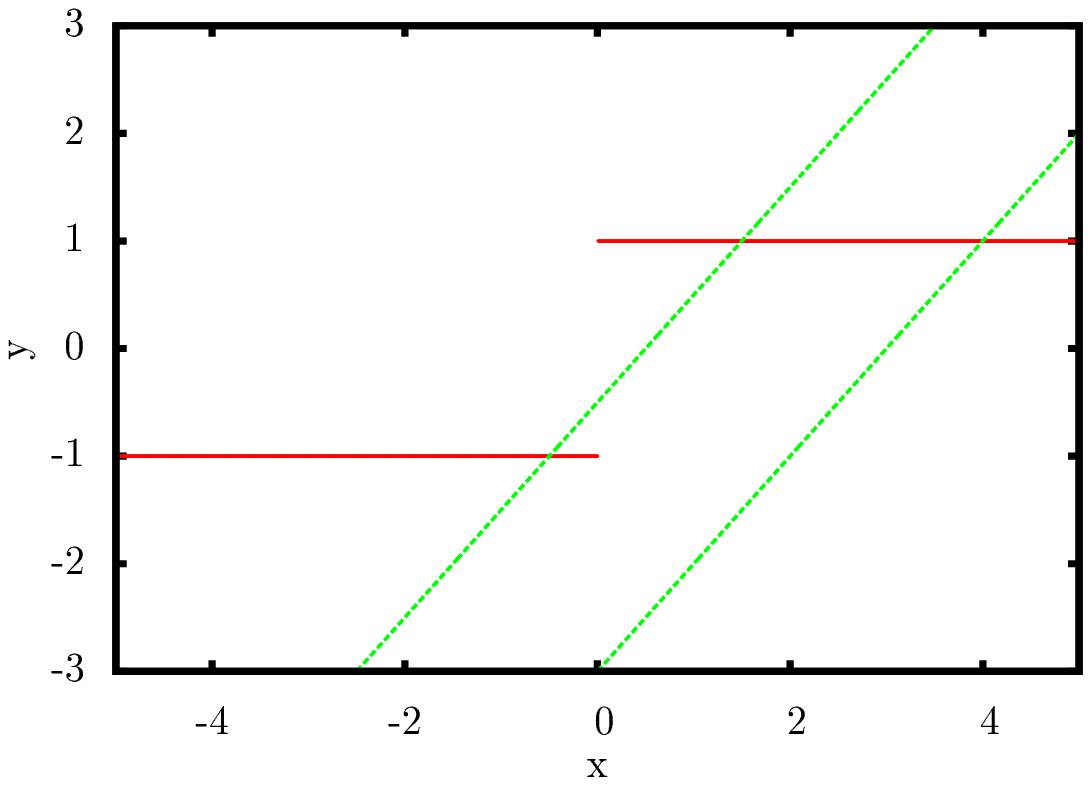}
\end{center}
\caption{
The solid (red) lines $y=1$ for $x>0$ and $y=-1$ for $x<0$ and the broken (green) lines $y=x-\hat{\phi}$, respectively, 
correspond to the left-hand side and the right-hand side of the lens equation (\ref{eq:lens_0}).
We plot the lines in the case $\hat{\phi}=0.5$ and $\hat{\phi}=3$.
The one or two intersections correspond to the real solutions of the lens equations.}
\label{fig:three}
\end{figure}
We obtain only one positive solution $\theta_{+}$ in the range $\hat{\phi} >1$ 
while we get a positive solution $\theta_{+}$ and a negative solution $\theta_{-}$ in the range $0\leq \hat{\phi} \leq 1$.

In the range $ \hat{\phi} >1$, by a straight forward calculation, we get 
\begin{eqnarray}
\frac{\hat{\theta}_{+}}{\hat{\phi}}\frac{d\hat{\theta}_{+}}{d\hat{\phi}}=1 + \frac{1}{\hat{\phi}}.
\end{eqnarray}
In the range $0\leq  \hat{\phi} \leq 1$, we obtain 
\begin{eqnarray}\label{eq:0_th_no_magnification_invariant}
\frac{\hat{\theta}_{\pm}}{\hat{\phi}}\frac{d\hat{\theta}_{\pm}}{d\hat{\phi}}=1\pm \frac{1}{\hat{\phi}}.
\end{eqnarray}
Therefore the signed magnification sum is $2$ in this range.

Only in the case $n=0$, 
the number of images is not always $2$.
The singular isothermal sphere lens is given by setting $n=0$ and $C=4\pi \sigma^{2}$, 
where $\sigma$ is the velocity dispersion of particles.

\section{III. ~signed magnification sums in the directly aligned limit}
In this section we will discuss the signed magnification sums of the general spherical lens model for $n\geqq 1$ 
in the directly aligned limit $(\hat{\phi} \sim 0)$.

For $\hat{\phi}>0$, 
the positive solution $\hat{\theta}_{+}(\hat{\phi})$ and the negative solution $\hat{\theta}_{-}(\hat{\phi})$ of the lens equation (\ref{eq:n+1th_lens_eq}) 
represent an outer image angle and an inner image angle
while $\hat{\theta}_{+}(\hat{\phi})$ and $\hat{\theta}_{-}(\hat{\phi})$ are an inner one and an outer one for $\hat{\phi}<0$, respectively.
The positive solution $\hat{\theta}_{+}$ monotonically increases as $\hat{\phi}$ increases.  
The signed magnifications of the images in the weak field limit are given by
\begin{eqnarray}
\mu_{0\pm}(\hat{\phi})
\equiv \frac{\hat{\theta}_{\pm}(\hat{\phi})}{\hat{\phi}} \frac{d\hat{\theta}_{\pm}}{d\hat{\phi}}( \hat{\phi} ). 
\end{eqnarray}

The lens equation (\ref{eq:n+1th_lens_eq}) has symmetry with respect to the point $\hat{\phi}=\hat{\theta}=0$,
so that 
\begin{eqnarray}
\hat{\theta}_{-}(\hat{\phi})=-\hat{\theta}_{+}(-\hat{\phi})
\end{eqnarray}
and
\begin{eqnarray}
\frac{d\hat{\theta}_{-}}{d\hat{\phi}} ( \hat{\phi} )
=\frac{d\hat{\theta}_{+}}{d\hat{\phi}} ( -\hat{\phi} ).
\end{eqnarray}
Thus, the relation of the magnifications is given by
\begin{eqnarray}
\mu_{0-}(\hat{\phi})
&=&\frac{\hat{\theta}_{-}(\hat{\phi})}{\hat{\phi}} \frac{d\hat{\theta}_{-}}{d\hat{\phi}}( \hat{\phi} ) \nonumber\\
&=&\frac{\hat{\theta}_{+}(-\hat{\phi})}{-\hat{\phi}} \frac{d\hat{\theta}_{+}}{d\hat{\phi}}( -\hat{\phi} ) \nonumber\\
&=&\mu_{0+}(-\hat{\phi}).
\end{eqnarray}

The positive image angle and the magnification in the directly aligned
limit ($\hat{\phi} \sim 0$) are given by 
\begin{eqnarray}
\hat{\theta}_{+}(\hat{\phi})
\sim 1+\frac{1}{1+n}\hat{\phi}+\frac{n}{2(1+n)^2}\hat{\phi}^2 
\end{eqnarray}
and
\begin{eqnarray}
\mu_{0+}(\hat{\phi})
\sim \frac{1}{1+n}  \frac{1+\hat{\phi}}{\hat{\phi}} ,
\end{eqnarray}
respectively.
From the symmetry, we can easily obtain the negative image angle and the signed magnification in the directly aligned limit
\begin{eqnarray}
\hat{\theta}_{-}(\hat{\phi})
\sim -1+\frac{1}{1+n}\hat{\phi}-\frac{n}{2(1+n)^2}\hat{\phi}^2 
\end{eqnarray}
and
\begin{eqnarray}
\mu_{0-}(\hat{\phi})
\sim -\frac{1}{1+n}  \frac{1-\hat{\phi}}{\hat{\phi}},
\end{eqnarray}
respectively.
Therefore the total magnification and the ratio of the magnifications in the directly aligned limit are given by
\begin{eqnarray}
\mu_{0}(\hat{\phi}) 
\equiv  \left| \mu_{0+}(\hat{\phi}) \right| + \left| \mu_{0-}(\hat{\phi}) \right| 
\sim \frac{2}{1+n}\frac{1}{\hat{\phi}}
\end{eqnarray}
and
\begin{eqnarray}
\left| \frac{\mu_{0+}(\hat{\phi})}{\mu_{0-}(\hat{\phi})} \right| \sim \frac{1+\hat{\phi}}{1-\hat{\phi}},
\end{eqnarray}
respectively.

Figure 4 shows that one can distinguish the general spherical lens models with their signed magnification sums $\mu_{0+}+\mu_{0-}$ which are less than unity.
We also can see that one can distinguish $n=1$ from $n=2$, $3$ and $4$ but one cannot distinguish between $n=2$, $3$ and $4$ for $\hat{\phi}\gtrsim  2$.
\begin{figure}[htbp]
\begin{center}
\includegraphics[width=80mm]{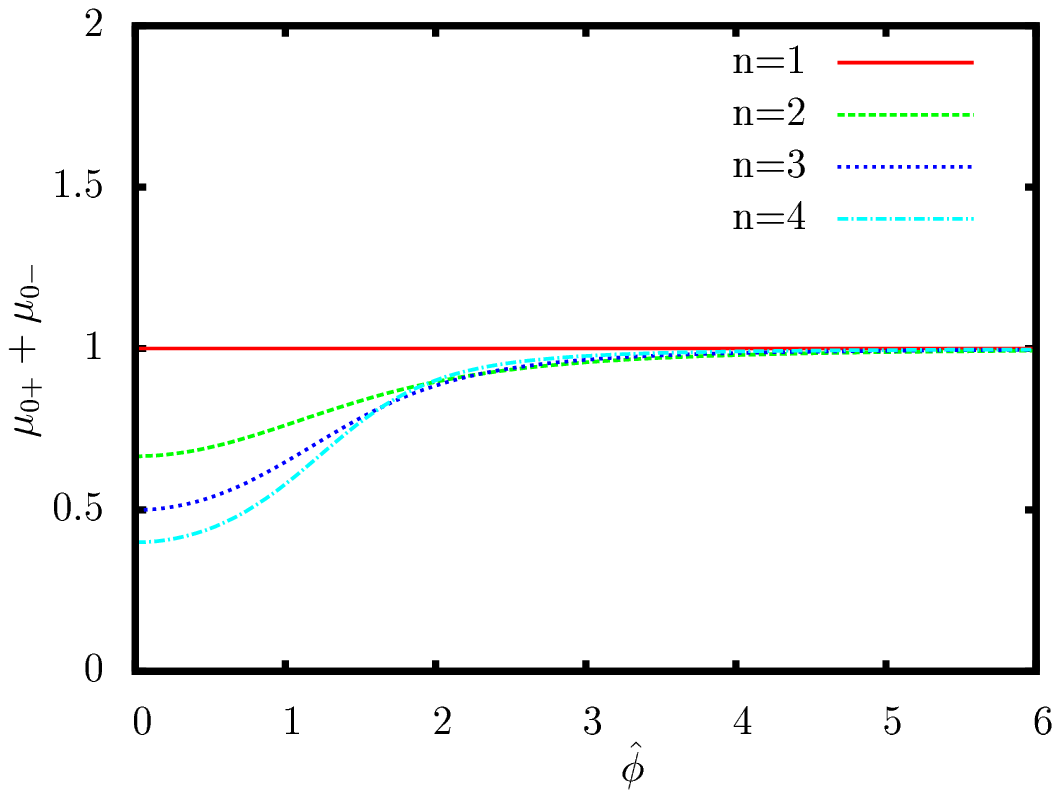}
\end{center}
\caption{The singed magnification sums of some general spherical lens models.
The solid, broken, dot and dot-dashed lines are the general spherical lens models for $n=1$, $2$, $3$ and $4$, respectively.
This shows that we can distinguish each models from the others.}
\label{fig:two}
\end{figure}
The minimum value of the signed magnification sums is given by 
\begin{eqnarray}
\lim_{\hat{\phi} \rightarrow 0} \left( \mu_{0+}(\hat{\phi})+\mu_{0-}(\hat{\phi}) \right)
=\frac{2}{1+n}.
\end{eqnarray}
The lower bound of the total magnification $\mu_{0}$ is given by 
\begin{eqnarray}
\frac{2}{1+n}
\leq \mu_{0+} + \mu_{0-}
\leq \left| \mu_{0+} \right| + \left| \mu_{0-} \right|
=\mu_{0}.
\end{eqnarray}
Therefore, gravitational lensing necessarily gives amplified light curves for $n=1$, while it does not necessarily for $n>1$.

From the lens equation~(\ref{eq:n+1th_lens_eq}), we obtain 
\begin{eqnarray}\label{eq:Jacobian}
\mu_{0\pm}
=\frac{\hat{\theta}^{2n+2}_{\pm}}{(\hat{\theta}^{n+1}_{\pm}\mp 1)(\hat{\theta}^{n+1}_{\pm}\pm n)}.
\end{eqnarray}
For $\hat{\phi}\gg 1$, Eq. (\ref{eq:Jacobian}) implies that $\mu_{0+}\simeq 1$ 
because $\theta_{+}\simeq \hat{\phi}\gg 1$, while $\mu_{0-}\ll 1$ 
because $\theta_{-}\ll 1$. In other words, if it is far from the alignment, the positive image is as luminous as the unlensed image, 
while the negative image is extremely faint.
Thus, we can ignore the gravitational lensing effects and the signed magnification sum $\mu_{0+}+\mu_{0-}$ becomes almost unity for $\hat{\phi} \gg 1$.

The difference of the reduced image angle in the directly aligned limit is given by
\begin{eqnarray}
\hat{\theta}_{+}-\hat{\theta}_{-} 
\sim 2+\frac{n}{(1+n)^2}\hat{\phi}^2.
\end{eqnarray}
Thus, the Einstein ring angle is given by
\begin{eqnarray}
\theta_{0}
\sim \frac{(1+n)^2(\theta_{+}-\theta_{-})}{2(1+n)^2+n\hat{\phi}^2}.
\end{eqnarray}

For $n=0$ these analyses are not valid in the region $1<\hat{\phi}$ because of the non-existence of the negative image angle $\hat{\theta}_{-}$.
However, it is valid in the region $0 \leq \hat{\phi} \leq 1$.
For $1<\hat{\phi}$, the magnification is
\begin{eqnarray}
1
\leq \mu_{0+}(\hat{\phi})
\leq 2.
\end{eqnarray}
and the total magnification $\mu_{0+}(\hat{\phi})+\mu_{0-}(\hat{\phi})$ is always 2 in the range $0 \leq \hat{\phi} \leq 1$.
So one can also distinguish the case $n=0$ from the other cases.

\section{IV. ~Deflection angle on the Ellis wormhole}
In this section, we briefly review the Ellis wormhole spacetime~\cite{Ellis_1973} and the deflection angle on it~\cite{Chetouani_Clement_1984,Tsukamoto_Harada_Yajima_2012,Nakajima_Asada_2012}.

The Ellis spacetime was investigated as a geodesically complete particle model by Ellis~\cite{Ellis_1973} and turned out to describe a wormhole connecting two 
Minkowski spacetimes. The Ellis wormhole spacetime is a static, spherically symmetric, asymptotically flat solution of the Einstein
equation with a massless scalar field with a wrong sign as a matter field.
Although such a matter field violates energy conditions, 
it could represent the negative energy density from the 
quantum effects, such as the Casimir effect. 
This spacetime is a typical and simplest example of wormholes 
proposed by Morris and Thorne~\cite{Morris_Thorne_1988,Morris_Thorne_Yurtsever_1988}.
This is a traversable wormhole in the sense that an observer 
can cross this wormhole in both directions. 

The line element in the Ellis wormhole solution is given by
\begin{equation}
ds^{2}=-dt^{2}+dr^{2}+(r^{2}+a^{2})(d\theta^{2}+\sin^{2}\theta d\phi^{2}),
\end{equation}
where $a$ is a positive constant corresponding the radius of 
the wormhole throat at $r=0$.
The photon is scattered if $|b|>a$, while reaches the throat if
$|b|\le a$. 
Since we are interested in the scattering problem, we assume $|b|>a$. 
Chetouani and Clement~\cite{Chetouani_Clement_1984} derived
the exact deflection angle $\alpha$ of light on the Ellis wormhole 
geometry as follows:
\begin{equation}
\alpha=2 K\left(\frac{a}{b}\right)-\pi,
\end{equation}
where $K$ is the complete elliptic integral of the first kind.
See e.g.~\cite{Handbook_of_Elliptic_Integrals_for_Engineers_and_Scientists}.
The deflection angle is diverging in the limit $|b|\to a$, while 
it is approximately given in the weak-field regime $|b|\gg a$ by
\begin{equation} \label{eq:weak_field_deflection_angle}
\alpha \approx \pm \frac{\pi}{4}\left(\frac{a}{b}\right)^{2}.
\end{equation}
Therefore, in our parametrization of general spherical lenses, 
the Ellis wormhole lens reduces to the case $n=2$ and
$C=\pi a^{2}/4$ in the weak-field regime.

\section{V. ~Discussion and conclusion}
It is well known that the signed magnification sum of the Schwarzschild lens is always unity in the weak field limit.
We realize that one can distinguish the exotic lenses with the parameter $n>1$ of the general spherical lens such as Ellis wormhole from mass lens systems
because the signed magnification sums of exotic lenses are less than unity.
It is also easy to determine $n$ by the signed magnification sums. 

The signed magnification sum is a powerful tool to find exotic lens objects 
because it only depends on the deduced source angle $\hat{\phi}$ and $n$
and we just have to observe the images for $\hat{\phi} \lesssim 1$ and for $\hat{\phi} \gg  1$ to determine the signed magnification sum.
However, we need a high resolution to observe the double images. 
We would also distinguish the lens objects with the ratio of magnifications of the double images and the total magnification. 
If we also measure the difference $\theta_{+}-\theta_{-}$ of the image angles, 
one can determine the Einstein ring angle $\theta_{0}$ and the source angle $\phi=\theta_{0}\hat{\phi}$.

Abe suggests that one can detect the Ellis wormholes by observing the light curves with the characteristic demagnification \cite{Abe_2010}.
Notice that the method to distinguish the lens objects with the signed magnification sums would be used in both the magnification and demagnification phases. 
Thus, we do not have to rely on only the demagnification to detect the Ellis wormholes.

Our method with the signed magnification sums is complementary to the methods to detect exotic lens objects 
with the light curves \cite{Abe_2010} and the astrometric image centroid displacements \cite{Toki_Kitamura_Asada_Abe_2011}.
To observe double images are much more feasible than to observe relativistic Einstein rings \cite{Tsukamoto_Harada_Yajima_2012} 
because relativistic images are faint and small and because relativistic rings are rare sights.

While the authors were finalizing the present paper, 
they noticed that Kitamura, Nakajima and Asada are taking the similar approach.
\\
\\
{\bf Acknowledgements:}
\\

The authors would like to thank 
H. Asada, T. Kitamura and K. Nakajima 
for valuable comments and discussion. 
The work of N. T. was supported in part
by Rikkyo University Special Fund for Research. 
T. H. was supported by the Grant-in-Aid for Young Scientists (B) (No. 21740190)
and the Grant-in-Aid for Challenging Exploratory Research (No. 23654082)
for Scientific
Research Fund of the Ministry of Education, Culture, Sports, Science
and Technology, Japan.
N. T. thanks the Yukawa Institute for Theoretical Physics at Kyoto University, 
where this work progressed during the YITP workshop YITP-W-12-08 on "Summer School on Astronomy \& Astrophysics 2012".

\end{document}